# LIQUID DENSITY SENSING USING RESONANT FLEXURAL PLATE WAVE DEVICE WITH SOL-GEL PZT THIN FILMS

*Jyh-Cheng Yu*[*] *and Huang-Yao Lin*

Department of Mechanical and Automation Engineering
National Kaohsiung First University of Science and Technology
2, Juoyue Rd., Nantz District, Kaohsiung 811, Taiwan, R.O.C.

**ABSTRACT**

This paper presents the design, fabrication and preliminary experimental results of a flexure plate wave (FPW) resonator using sol-gel derived lead zirconate titanates (PZT) thin films. The resonator adopts a two-port structure with reflecting grates on the composite membrane of PZT and $SiN_x$. The design of the reflecting grating is derived from a SAW resonator model using COM theory to produce a sharp resonant peak. The comparison between the mass and the viscosity effects from the theoretical expression illustrates the applications and the constraints of the proposed device in liquid sensing. Multiple coatings of sol-gel derived PZT films are adopted because of the cost advantage and the high electromechanical coupling effect over other piezoelectric films. The fabrication issues of the proposed material structure are addressed. Theoretical estimations of the mass and the viscosity effects are compared with the experimental results. The resonant frequency has a good linear correlation with the density of low viscosity liquids, which demonstrate the feasibility of the proposed device.

## 1. INTRODUCTION

Acoustic wave devices have attracted enormous attention for sensor application because that the wave velocity and damping are sensitive to external disturbances such as temperature, pressure, additive mass, and viscosity [1]. Among them, the phase velocity of lamb waves, unlike the velocity of surface acoustic wave (SAW), depends on the thickness of the propagating plate. They can be considered as two Rayleigh waves that are strongly coupled and propagating on both sides of the plate. Two vibrating modes can propagate through the plate independently, namely the symmetric and the anti-symmetric lamb modes. The anti-symmetric zero mode $A_0$, also called "flexural plate wave" (FPW), propagating on a thin membrane with thickness 5% or less of the acoustic wavelength, has a phase velocity lower than the sound velocity of loading liquid. FPW sensors are suitable for liquid sensing because a slow mode of propagation, such as $A_0$ mode, will minimize the radiation energy loss.

The excitation and the detection of the acoustic waves are most readily accomplished by the use of interdigital transducers (IDTs) [2] on thin piezoelectric plate that is used to be realized from the etching processes on a bulk piezoelectric substrate. Piezoelectric thin films have the cost advantage over crystalline materials. Many literatures addressed FPW sensors using ZnO[3] and AlN[5]. The electromechanical coupling effect and the dielectric constant of PZT are much higher than AlN and ZnO, which makes PZT films potentially suitable for sensor application. However, the polycrystalline structure of PZT and the required high temperature of heat treatment during the coating process complicate the realization.

The potential applications of FPW sensors to chemical and liquid sensing have attracted a lot of research interests. Costello *et al.*[3] proposed a simple theory for the mass sensitivity of a delay-line oscillator with ZnO on silicon nitride membrane, and modeled the attenuation of plate waves in contact with viscous liquids. Laurent *et al.*[4] addressed the configuration design of the FPW devices using AlN and ZnO on silicon membrane, and showed that the FPW device has a large mass sensitivity compared to other acoustic devices. Weinberg *et al.*[5] derived the fluid-damping model for resonant FPW devices. To increase the differentiability of the resonant frequency shift, reflecting gratings are added to the FPW devices that is first reported by Joshi[6] using a Y-X lithium niobate plate. Nakagwa[7] also adopted the same configuration but on an AT-cut quartz substrate.

Mass, tensile stress, and viscosity effects might couple when in contact with liquids, which will complicate the differentiablity in sensor applications. Very few literatures address the practical issues and strategy in liquid density sensing.

This study will discuss the design and application issues for the FPW resonator using the sol-gel derived PZT on silicon nitride membrane as shown in Figure 1. We will derive the design of the reflecting grating using the Coupling of Modes (COM) theory, and apply to the FPW devices. The mass loading when the device is in contact with liquid will introduce the deviation of resonant frequency that is determined by the liquid density and viscosity. The constraints of the proposed device in liquid density sensing will be discussed. Finally, experimental result is compared with the theoretical and the difference is discussed.

## 2. DESIGN OF REFLECTORS

The design of the reflecting grating is derived from a two-port SAW resonator over a bulk PZT modeled using the COM theory[9]. There are three basic elements in the modeling of a typical SAW resonator: IDT, spacing, and reflector that can be described by three complex transmission matrices of [T], [D], and [G] respectively as shown in Figure 2. Matrix [T] is a 3×3 transmission matrix for the IDTs, including both the acoustic and the electric parameters. The terms, *a* and *b,* represent the incident and the reflected electrical signals. Matrix [D] is a 2×2 matrix for the acoustic transmission line between IDTs and gratings. Matrix [G] is a 2×2 matrix for the SAW reflection grating to describe the relationship between the acoustic transmission and reflection responses $W_i$.

If there is no input electrical signal for the output IDT, the overall acoustic modeling can be simplified as follows:

$$[W_0] = [M][W_7] + a_3[G_1][D_2][\tau_3] \quad (1)$$

where $[M] = [G_1][D_2][T_3][D_4][T_5][D_6][G_7]$

$\tau_3$ is the column matrix relating to the input coupling.

Substitute the required boundary conditions, and we can obtain the simulation of the transmission coefficient $S_{21}$.

The simulation parameters used here are as follows: acoustic wavelength is 40 (μm) with uniform finger spacing, separation between IDTs = 10 wavelengths, and the phase velocity of bulk PZT is 2400 (m/s). The reflection phase, *θ,* is determined by the position of standing wave at the reference plane relating the sign of reflected-to-incident surface waves entering the reflection grating, which will affect the spacing design between the gratings and adjacent IDTs ($D_2$ and $D_6$). This determination of reflecting phase is difficult and often depends on the experimental for the material combination of the piezoelectric substrate and the reflecting electrodes[9]. The reflection coefficient for the PZT thin film and Pt/Ti reflecting grate is not available. However, since PZT is a strong piezoelectric, like lithium niobate, the reference phase $\theta = 0°$ is assumed for the open circuit design of reflector.

Based on the above assumptions and the tradeoff between the device size and performance, we then determine the following design parameters.
(1) The number of reflection grating is set 40 that are to form standing waves to reduce insertion loss.
(2) The more pairs of IDT, the smaller insertion loss and the wider bandwidth. The number of pairs of IDT is selected as 20.
(3) The overlap length of IDT is 50λ because it has a low insertion loss and high transmission effect, where λ is the acoustic wavelength.

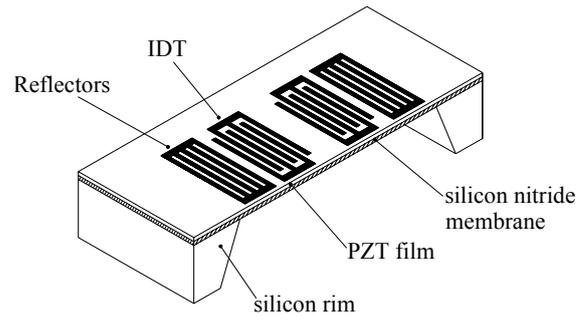

Figure 1 Schematic View of FPW resonators

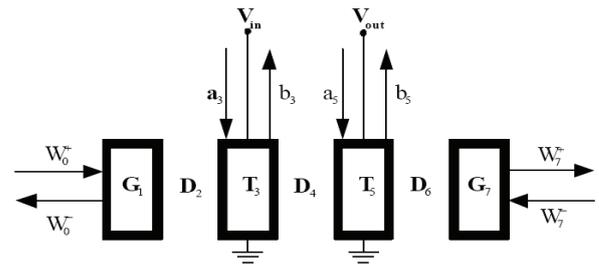

Figure 2 Representation of two-port resonator building blocks

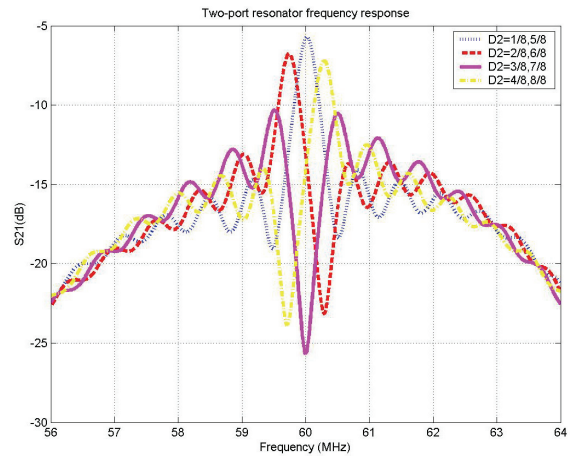

Figure 3 Two-port SAW resonator frequency response

(4) The separation between the IDTs is 10λ. The spacing between the gratings and adjacent IDTs (*D2*) are (1/8+n/2)λ to produce a sharp resonant peak (Figure 3).

The derived reflector design will be applied to the FPW device to increase the differentiablity of frequency deviation due to liquid loading.

## 3. THE SENSING MECHANISM OF FPW DEVICES

### 3.1. Estimation of phase velocity

Consider the $A_0$ mode of FPW propagating on a thin plate; the phase velocity of the plate regime can be expressed by a simple asymptotic expression [8]:

$$v_P = \left(\frac{B}{M}\right)^{1/2} \quad (2)$$

where $B$ is the bending stiffness and $M$ is the mass per unit area of a homogeneous isotropic plate.

When the device is in contact with liquid, it will introduce additional stiffness effect due to liquid weight and additional mass effect due to the agitation of the liquid. The phase velocity of the plate regime subject to a tensile stress and liquid loading can be well approximated as follows [8]:

$$v_P = \left(\frac{T_x + B}{M + \rho_F \delta_E + M_\eta}\right)^{1/2} \quad (3)$$

where $T_x$ is the component of in-plane tension in the $x$ direction,
$\rho_F \delta_E$ is the mass effect,
$\rho_F$ is the density of the fluid, and
$M_\eta$ is the viscosity effect.

$$\delta_E \approx \left(\frac{\lambda}{2\pi}\right) \quad (4)$$

if the phase velocity is much less than the speed of sound in the contact liquid.

$$M_\eta = \frac{\rho_F \delta_V}{2} \quad (5)$$

where $\delta_V = \left(\frac{2\eta}{\omega \rho_F}\right)^{1/2}$ is the viscous decay length,

$\omega$ is the operating angular frequency, and
$\eta$ is the shear viscosity.

The material structure of the proposed FPW device is Pt (0.15μm)/Ti (0.02μm)/PZT(1μm)/LSMO (0.1μm)/SiN$_x$ (1.2μm). Table 1 lists mechanical properties of the material system. The PZT properties are assumed from bulk material. The (La$_x$Sr$_{1-x}$)MnO$_3$, LSMO, is a buffer layer between PZT and SiN$_x$. The material properties of LSMO are not available and assumed to be the same as the PZT film. The mass effect of the Pt and the Ti are assumed negligible.

Table 1 The material properties of the composite plate

|  | SiN$_x$ | PZT + LSMO | Ref. |
|---|---|---|---|
| Thickness (μm) | 1.2 | 1 PZT+0.1 LSMO | |
| Young's modulus ($E$, N/m$^2$) | 3.85*10$^{11}$ | 8.6*10$^{10}$ | [10] |
| Poisson ratio ($v$) | 0.27 | 0.25 | |
| Density ($\rho$, kg/m$^3$) | 3100 | 7600 | |
| $M$ (kg/m$^2$) | 0.00372 | 0.00836 | |

The thickness of the composite plate is 2.3 (μm). The material parameters of the membrane as shown in Table 2 can be estimated from the composite plate theory. The estimated phase velocity is 235.06 (m/s) from Eq.(3), and the resonant frequency of the device in air will be $f_{air}$ = 5.88 (MHz).

Table 2 The parameters of the composite plate

| $E$ | 2.42*10$^{11}$ (N/m$^2$) |
|---|---|
| $M$ | 0.1176 (N/m$^2$) |
| $v$ | 0.26 |
| $E'$ | 2.6*10$^{11}$ (N/m$^2$) |
| $B$ | 6497.93 (N/m) |

### 3.2. The loading effects for a low viscosity liquid

When the FPW device is in contact with a low viscosity liquid, the viscosity effect of eq. 5 is negligible and the phase velocity will be influenced by the mass effect ($\rho_F \delta_E$) that can be determined by the evanescent decay length. If the loading liquid is a small droplet on the thin plate and the contacting surface is not hydrophilic, the droplet will not spread out evenly and remain hemispherical as shown in Figure 4. Because of the evanescent decay length is approximate $\lambda/2\pi$, not all the mass of the loading drop contributes the mass loading effect. Because the shape of the drop can't be accurately controlled, the change of the phase velocity usually is not in proportional to the number of liquid droplets.

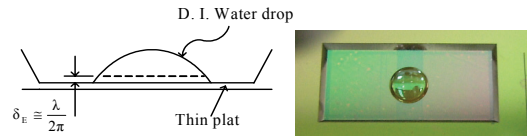

Figure 4 The liquid droplet on the FPW device

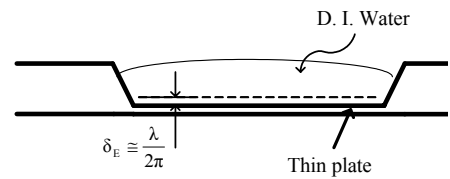

Figure 5 The evanescent decay length for the liquid loading in a FPW device

Therefore, the liquid is suggested to fill up the cavity if the device is used to detect the liquid density. The liquid density will determine the low viscosity liquid loading if the liquid level is higher than the evanescent decay length as shown in Figure 5.

The liquid weight will introduce the tensile stress in the membrane that will result in the deviation of phase velocity. The mass sensitivity and tension sensitivity of the perturbation of phase velocity are as follows:

$$\frac{\Delta v_p}{v_p} = s_m \times \rho_F + s_T \times T_x \quad (6)$$

where $s_m = -\frac{\delta_E}{2(M + \rho_F \delta_E)}$ and $s_T = \frac{1}{2(T_x + B)}$

The calculated phase velocity for the composite membrane in air is about 235 (m·s$^{-1}$) which is much smaller than the speed of sound in water, 1482 (m·s$^{-1}$). The stiffness of the membrane is estimated using the composite plate theory. The calculated sensitivities in our case are $s_m$ =-2.78(m$^2$/N) and $s_T$ =7.69*10$^{-5}$(m/N). The acoustic wavelength is 40 (μm). Assuming 5 (mg) of water is loaded on the cavity, the average tensile stress along the wave propagating direction can be estimated from a finite element analysis. The estimated frequency deviation due to the mass loading of water is about -0.94 (MHz) and the frequency deviation due to the tensile effect is only +1.24 (Khz). The tension effect due to the liquid pressure can be ignored when compared with the mass loading effect. Therefore, when the device cavity is filled up with different liquids, the resonant frequency deviation will relate to the liquid density, and thus can be used as a density sensor.

We can obtain the phase velocity for the FPW device loaded with three low viscosity liquids as shown in Table 3.

Table 3 Estimated resonant frequency of the FPW device loaded with low viscosity liquids

|  | IPA | Water | Saline solution |
|---|---|---|---|
| Viscosity N-s/m$^2$ | 0.0025 | 0.001 | ~0.0015 |
| Phase Velocity (m/s) | 197.48 | 190.05 | 183.77 |
| Resonant Frequency (MHz) | 4.94 | 4.75 | 4.59 |

### 3.3. The loading effects for viscous liquid

From Eq.(3), we notice that the phase velocity of FPW will be influenced by fluid density ($\rho_F$) and shear viscosity ($\eta$) of the loading liquid. However, we can't differentiate the velocity perturbations between liquid density and viscosity because the density and viscosity are coupled in the viscous effect as seen from Eq. (5). In another word, the liquid viscosity can't be determined by the frequency shift. Therefore, the proposed device is not suitable for the density sensing of viscous liquids and the viscosity sensing via the resonant frequency deviation.

### 4. FABRICATION PROCEDURE

The schematic view of the FPW resonator is shown in Figure 1. The materials system of the resonator is assumed Pt/Ti/PZT/LSMO/SiN$_x$. The LSMO and the PZT thin films are multiple-coated by sol-gel techniques. Furnace heating of 650ºC is used to transform the thin film into polycrystalline piezoelectric layer. Higher heat treatment temperature might improve the material characteristics of PZT, severe thermal stresses may crack the film due to incompatibility among constituent layers of the composite membrane. The LSMO is used as a buffer layer between PZT and SiN$_x$, which can enhance the piezoelectric characteristic and avoid possible cracking of the PZT. As seen from the comparison between Figure 6 and Figure 7, the LSMO buffer layer not only enhances the Residual Polarization but also decrease the polarization decay due to fatigue loading that is very important for resonant devices. The electrodes of the IDT are patterned for the period of 40 (μm) using lift-off. Finally, the membrane cavity of the device is fabricated using KOH anisotropic etching (30ºC, 80%). The final composite membrane is consisted of 1.2 (μm) silicon nitride and 1.1 (μm) PZT layer. The size of the rectangular membrane is about 4.2×2.7(mm).

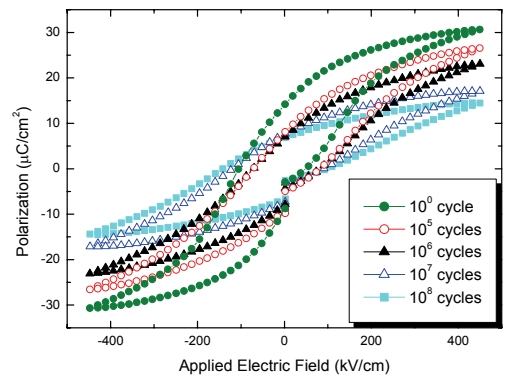

Figure 6 The fatigue characteristics for the PZT film without LSMO buffer layer

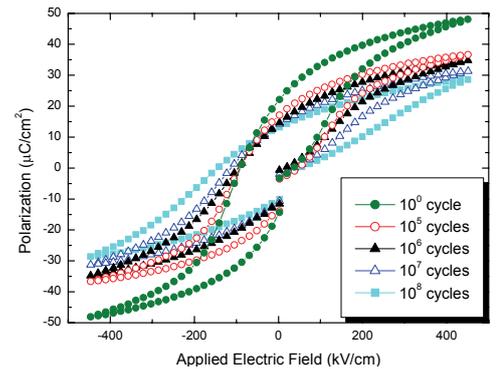

Figure 7 The fatigue characteristics for the PZT film with LSMO buffer layer

## 5. EXPERIMENTAL RESULTS AND DISCUSSIONS

### 5.1. The FPW delay line and the FPW resonator signals

The frequency responses of the FPW delay line and the FPW resonator are measured using a network analyzer (HP8753 ES). The frequency responses of the FPW delay line and the FPW resonator are shown in Figure 8. The measured resonant frequency without liquid loading is 5.53 (MHz) that is pretty close to the theoretical estimation of 5.88(MHz). The difference might be due to the use of the bulk material properties in the theoretical estimation because the PZT film properties are not available. The resonator has a more obvious resonant signal than the delay line design, although sharp peak is not observed, which might be due to the polycrystalline structure of PZT. Preliminary poling of the piezoelectric film shows improvement but further investigation is required.

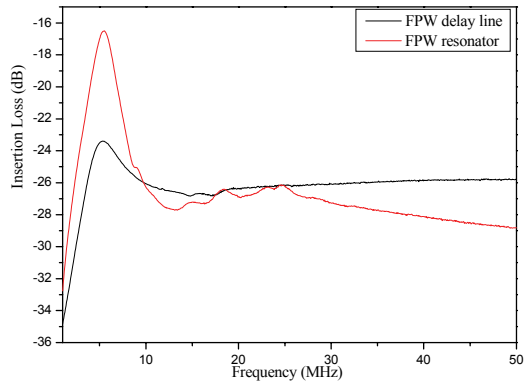

Figure 8 The $S_{21}$ frequency response of the FPW devices

### 5.2. Liquid sensing using the FPW resonator

Three low viscosity liquids, DI Water (1 g/cm$^3$), IPA (0.787 g/cm$^3$) and saline solution (1.2 g/cm$^3$), are applied to the resonator. The viscosity effect is negligible compared with the mass effect. The sensitivity analysis between the resonant frequency and the densities of the liquids is summarized in Figure 9. The theoretical relative density sensitivity for low viscosity liquids is -0.848 (Mhz/g·cm$^{-3}$). The results show that the resonant frequency and the liquid density have a good linear correlation despite a static difference, which demonstrates the feasibility of density sensing. The static difference may be due to liquid damping and stress effects that are presumed negligible.

However, the phase velocity of FPW will be influenced by fluid density and shear viscosity of the high viscosity liquid from the theoretical analysis. We can't differentiate the velocity perturbations between liquid density and viscosity because the density and viscosity are coupled in the viscous effect. In another word, the liquid viscosity can't be determined by the frequency shift. Here, we compare two liquids, glycerol with viscosity of 934×10$^{-3}$ (N-s/m$^2$) and a low viscosity saline solution with the same density (1.2 g/cm$^3$) to study the viscosity effect. From Table 4, we observe that the glycerol loading will introduce additional frequency deviation compared with the saline solution with the same density. Also, the viscosity effect of glycerol will cause more damping effect than the saline solution and increases the insertion loss.

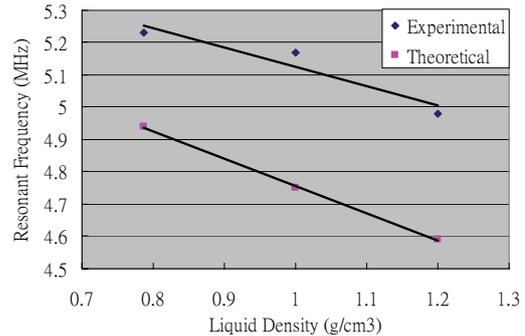

Figure 9 The sensitivity analysis between the resonant frequency and the density for low viscosity liquids

Table 4 The viscosity effect on the resonant frequency and insertion loss

|  | Theoretical | Experimental | |
|---|---|---|---|
|  | Frequency (MHz) | Frequency (MHz) | Insertion Loss (dB) |
| Saline solution | 4.59 | 4.98 | -33.38 |
| Glycerol | 4.49 | 4.73 | -37.04 |

## 6. CONCLUSIONS

This study has successfully fabricated the FPW resonator on the PZT piezoelectric thin films and compared with the modeling analysis. Although the polycrystalline structure of PZT may obscure the resonant peak of the frequency response, it is good enough to differentiate the frequency deviation. The device is applied to the loading of various liquids. We have observed that the liquid loading will increase the insertion loss and decrease the resonant frequency that is consistent with theoretical prediction. The linear correlation between the resonant frequency and the density of low viscosity liquid demonstrates the feasibility of density sensing. Additional frequency deviation insertion loss are observed for high viscosity liquids, which also matches fairly with the theoretical estimation. However, the proposed device can't differentiate the velocity perturbations between liquid density and viscosity, which prevent the application of the device in the density sensing of high viscosity liquid via frequency deviation alone. The accompanying increase of the insertion loss due to liquid viscosity may be applied together to determine the liquid.